\documentclass[amssymb,prc,floatfix,amsmath,twocolumn,reprint]{revtex4-1}
\usepackage{graphicx,bm}
\usepackage{bm,color}
\usepackage{epstopdf}
\usepackage{natbib}
\usepackage{array,tabu}
\newcommand{\etal}{{\em et al.}}
\newcommand{\A}[1]{$^{#1}$}
\newcommand{\vlk}{V_{\text{low-}k}}

\newcommand{\epinu}{$(e_{\pi},e_{\nu})$\,}
\newcommand{\enupi}{$(e_{\nu},e_{\pi})$\,}
\newcommand{\eff}{e$^2$fm$^4$\,}
\newcommand{\bet}{$B(E2: 2^+\to 0^+)$\,}
\newcommand{\bef}{$B(E2: 4^+\to 2^+)$\,}
\newcommand{\beq}{$B(E2: 4^+\to 2^+)/B(E2: 2^+\to 0^+)$\,}
\newcommand{\betm}{B(E2: 2^+\to 0^+)}
\newcommand{\bl}{\begin{Large}}
\newcommand{\el}{\end{Large}}
\newcommand{\be}{\begin{equation}}  
\newcommand{\ee}{\end{equation}}
\newcommand{\bg}{\begin{gather}}  
\newcommand{\eg}{\end{gather}}
\newcommand{\ba}{\begin{eqnarray*}}
\newcommand{\ea}{\end{eqnarray*}}

\newcommand{\hw}{$\hbar \omega \,$}
\newcommand{\hbw}{\hbar \omega}

\newcommand{\ie}{{\it i.e.,\ }}
\newcommand{\q}{$\langle 2q_{20}\rangle\,$}

\def\tcn{\textcolor{black}}
\begin{document}
\title{Quadrupole dominance in light Cd and Sn isotopes }
\author{ A.~P.~Zuker}
\affiliation{Universit\'e de Strasbourg, IPHC, CNRS, UMR7178 Strasbourg, France}

\email{andres.zuker@in2p3.cnrs.fr}
\begin{abstract}
  Shell model calculations with the neutron effective charge as single
  free parameter describe well the \bet and \bef rates for $N\le 64$
  in the Cd and Sn isotopes. The former exhibit weak permanent
  deformation corroborating the prediction of a pseudo SU3 symmetry,
  which remains of heuristic value in the latter, though the pairing
  force erodes the quadrupole dominance. Calculations in $10^7$ and
  $10^{10}$-dimensional spaces exhibit almost identical patterns: A
  vindication of the shell model. For $N\ge 64$ quadrupole dominance
  is accentuated in the Cd isotopes and gives way to seniority dominance
  for the Sn isotopes. 
%
\end{abstract}
\date{\today} \maketitle 

\begin{figure}[h]
\begin{center}
\includegraphics[width=0.5\textwidth]{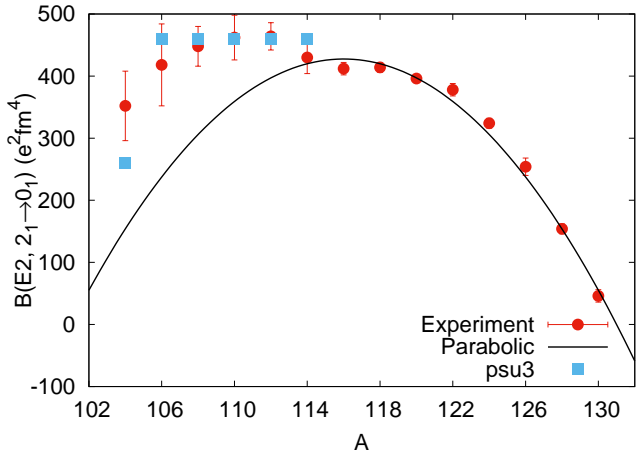}
\end{center}
\caption{\label{fig:SnBE2} The experimental \bet for the Sn isotopes
  from compilations~\cite{be22016}, compared with some arbitrary
  parabolic shape and pseudo-SU3 results to be explained here
  (squares).}
\end{figure}

All nuclear species are equal, but some are more equal than
others. The tin isotopes deserve pride of place, because $Z=50$ is the
most resilient of the magic numbers, because they are very numerous,
and many of them stable, starting at $A=112$. For these, accurate data
have been available for a long time. As seen in Fig.~\ref{fig:SnBE2} a
parabola accounts very well for the \bet trend, except at
\A{112-114}Sn.

That these early results (Jonsson \etal~\cite{JONSSON1981}) truly
signal a change of regime became evident through work on the unstable
isomers, starting with the measure in \A{108}Sn by Banu \etal
~\cite{banu2005}. A flurry of activity followed~\cite{ vaman2007sn,
  cederkall2007sub, ekstrom2008sn, kumar2010enhanced,
  bader2013quadrupole, doornenbal2014intermediate,
  kumar2017noevidence}, from which a new trend emerged in which the
parabola---characteristic of a seniority scheme-----gives way to a
platform, predicted by a pseudo SU3 scheme (the squares).  Here we are
going a bit fast to follow the injunction of Montaigne: start at the
end (``Je veux qu'on commence par le dernier poinct'' Essais II
10)~\cite[p. 298]{montaigne}. To slow down, we go back to the origin of
this study, the Cadmium isotopes, where quadrupole dominance is
stronger and its consequences more clear-cut.

\begin{figure}[ht]
\begin{center}
\includegraphics[width=0.5\textwidth]{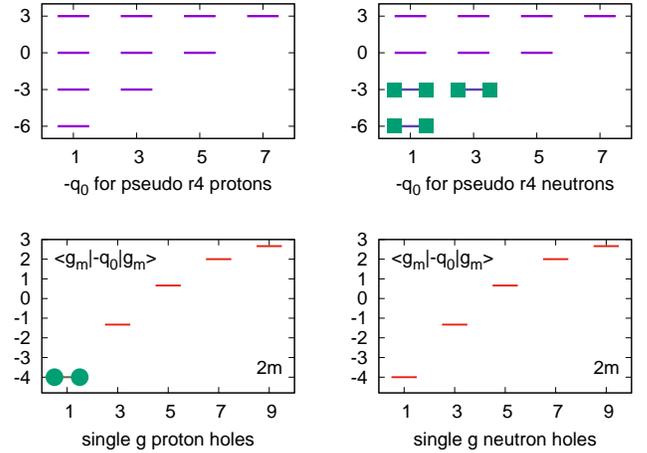}
\end{center}
\caption{\label{fig:SP} The SP spaces adapted to the Cd and Sn
  isotopes. The (dimensionless) $q_0=$\q values correspond to single
  particle and hole occupancies for the pseudo $r_4$ and $g$ cases
  respectively. The minus sign is an artifact to make occupancies
  start from the bottom. The figure illustrates the \A{104}Cd
  configration: circles for holes and squares for particles.}
\end{figure}
\section{The Cd isotopes}\label{sec:cd}

The basic idea is inspired by Elliott's SU3
scheme~\cite{elliotta,elliottb} and consists in building intrinsic
determinantal states that maximize $q_0$, the expectation value of the
quadrupole operator $\hat q_0=2q_{20}$, \ie
$q_0$=\q~\cite{Q,rmp,nilssonSU3}.  Fig.~\ref{fig:SP} implements the
idea for \A{104}Cd ($Z=48,\, N=56$).  The single shell (S)
contribution of the $g_{9/2}\equiv g$ proton orbit (S$g$) is given
by~Eq.\eqref{q0S} (with changed sign for hole states).  For the
neutron orbits, the pseudo SU3 scheme
~\cite{arimapsu3,hecht1969,nilssonSU3} (P generically, P$r_p$ for
specific cases) amounts to assimilate all the orbits of a major
osillator shell of principal quantum number $p$, except the largest
(the $r_p$ set) to orbits in the $p-1$ major shell. In our case the
$sdg$ shell has $p=4$, and $r_4$ is assimilated to a $pf$ shell. As
the $\hat{q_0}$ operator is diagonal in the oscillator quanta
representation, maximum \q is obtained by orderly filling states
$(n_z\, n_y\, n_x=(300),(210),(201)\ldots (012),(003)$, with
$q_0=$\q=$2n_z-n_y-n_z=$ 6, 3, 3, 0, 0...-3, -3, as in
Fig.~\ref{fig:SP}. Using $q(n)$ for the cumulated $q_0$ value (e.g. 24
for \A{104}Cd in Fig.~\ref{fig:SP}), the intrinsic quadrupole moment
then follows as a sum of the single shell (S) and pseudo SU3 (P)
contributions

\begin{gather} 
q_0(S)=2\langle r^2C_{20}\rangle =\sum_m(p+3/2)\frac{j(j+1)-3m^2}{2j(j+1)} \label{q0S}\\
q_0(P)=q(n),~~~Q_0(SP)=[(8e_{\pi}+q(n)e_{\nu})b^2]~{\rm e}{\rm fm}^2\label{Q0SP}
\end{gather}

where we have introduced effective charges \epinu for protons and
neutrons, and recovered dimensions through $Q=b^2q$ with

$ b^2\approx 41.4/\hbw~ {\rm fm}^2$ and $\hbw=45A^{-1/3}-25A^{-2/3}\label{hbw}$

To qualify as a Bohr Mottelson rotor, $Q_0(SP)$ must coincide with the
intrisic spectroscopic $Q_{0s}$ and transition  $Q_{0t}$ quadrupole
moments, defined through (as, e.g, in Ref.~\cite{nilssonSU3})

\begin{gather}
Q_{spec}(J)=<JJ\vert3z^2-r^2\vert
JJ> \nonumber \\ 
Q_{0s}=\frac{(J+1)\,(2J+3)}{3K^2-J(J+1)}\,Q_{spec}(J), \quad K\ne1  
\label{bmq}\\
 B(E2,J\rightarrow J-2)=
 \frac{5}{16\pi}\,e^2|\langle 
JK20| J-2,K\rangle |^2 \, Q_{0t}^2\label{bme2}\\
K\ne 1/2,\, 1,\betm=Q_{0SP}^2/50.3 ~{\rm e}^2{\rm fm}^4\label{be2} 
\end{gather}

To speak of deformed nuclei two conditions must be met: \beq=1.43
(the Alaga rule from Eq.~\eqref{bme2}), and the ``quadrupole
quotient'' rule, $Q/q$ which follows from Eqs.\eqref{bmq} and
\eqref{bme2} by equating $Q_{0s}\approx Q_{0t}$:

\begin{gather}\label{qQ}
50.27B(E2:2^+\to
0^+)/(3.5Q_{spec})^2=(Q/q)^2\approx 1
\end{gather}

Full verification demands calculations but
Eq.~\eqref{be2} can be checked directly by inspecting 
Fig.~\ref{fig:SP} as done in Table~\ref{tab:SP}.
\begin{table}[h]
  \caption{\label{tab:SP} \bet estimates for \A{98+n}Cd in \eff from
    Eq.\eqref{be2}. B20sp uses
    naive $q(n)_n$ from diagonalization of $\hat q_0$ in the $pf$ shell
    \ie strict SU3, with $(e_{\nu},e_{\pi})=(1.1,1.7)$.  The B20SP
    numbers use (full) $q(n)_f$ from diagonalization of $\hat q_0$ in the $r_4$
    space, $(e_{\nu},e_{\pi})=(1.0,1.5)$. The $b^2$ values
    range from 4.78 fm$^2$ for $A=98$ to 4.94 fm$^2$
    for $A=110$.  Experimental values (B20e) for \A{102-104}Cd are taken
    from~\cite{Cd100-104} and \cite{Cd102-104}, and from
    compilations~\cite{be22016} for \A{106-110}Cd.}
\begin{tabular*}{\linewidth}{@{\extracolsep{\fill}}|c|cccccc|}
\hline
$A$&100&102&104&106&108&110\\
$n$&2&4&6&8&10&12\\
$q(n)_n$&12&18&24&24&24&24\\
$q(n)_f$&14.8&22.6&29.5&30.0&29.6&29.3\\
B20e&$<$560(4)&562(46))&779(80)&814(24)&838(28)&852(42)\\
B20sp&330 &517&751&756&770&776\\  
B20SP&330&555&809&838&833&827\\
\hline
\end{tabular*}
\end{table}

Note that the naive form of P used so far (in $q(n)_n$ and B20sp) is
supplemented by the more accurate $q(n)_f$ and B20SP using fully
diagonalized values of \q.  The remarkable property of the $r_4^n$
space, that produces four identical $q(n)_s$ values for $m=6-12$, has
already been put to good use in Ref.~\cite{Q} and Ref.~\cite[Fig. 38, Table
VII]{rmp}. In the present case it is seen to do equally well.

Now for the shell model diagonalizations in spaces defined by
$(g^{X-u}{r}_{4}^u)_{\pi}(g^{10-t}r_4^{n+t})_{\nu}$, $X=8$ for Cd and
10 for Sn. The proton ($u$) and neutron ($t$) excitations are
restricted to have $u+t \leq M$.  The calculations were done for
$utM=000$ (the case in Fig.~\ref{fig:SP}), 111, 101 and 202 using
$\vlk$ variants~\cite{vlk} of the precision interaction
N3LO~\cite{N3LOa} (denoted as I in what follows) with oscillator
parameter \hw= 8.4 MeV and cutoff $\lambda=2$ fm$^{-1}$.  As a first
step the monopole part of I is removed and replaced by single-particle
energies for $^{100}$Sn from Ref.~\cite{gemo}(\tcn{referred to as GEMO
  for General Monopole: a successful description of particle and hole
  spectra on magic nuclei from Oxigen to Lead, in particular
  consistent with the analysis of Ref.~\cite{Sn100} for \A{100} Sn.})

 The I interaction is then subject to an overall 1.1 scaling and
renormalized by increasing the $\lambda \mu=20$ quadrupole and $JT=01$
pairing components by q$\times$10\% and p$\times$10\%, respectively.
The resulting interactions are called I.q.p.  According to
Ref.~\cite{mdz} the quadrupole renormalization (due to 2\hw
perturbative couplings) amounts to 30\%, a theoretically sound result,
empirically validated by the best phenomenological interactions in the
$sd$ and $pf$ shells.  By the same token the effective charges in 0\hw
spaces are estimated as \enupi= (0.46, 1.31), as confirmed in
Refs.~\cite{BE2FeCr, PhysRevC.69.064304}.  For the pairing component,
perturbation theory is not a good guide, but comparison with the
phenomenological interactions demands a 40\% increase~\cite{mdz,rmp}.
It follows that I.3.4 and \enupi= (0.46, 1.31) should be taken as
standard for full 0\hw spaces.

As we will be working in very truncated ones, renormalizations should
be implemented. A hint comes from the need to reduce the very large
effective charges invoked in Table~\ref{tab:SP} through polarization
mechanisms that involve excitations to the $g$ shell. Proton jumps
will contribute to $e_{\nu}$ and are expected to have greater impact
than the corresponding neutron jumps, rapidly blocked by the
$(r_4^{n+t})_{\nu}$ particles. As a consequence we set $e_{\pi}=1.4$,
a guess close to the standard value, and let $e_{\nu}$ vary, thus
becoming the only adjustable parameter in the calculations. A choice
validated later in section~\ref{sec:SM}.

\begin{figure}[ht]
  \begin{center}
    \includegraphics[width=0.4\textwidth]{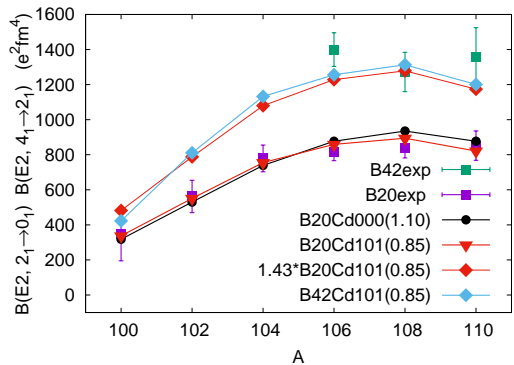}
\end{center}
\caption{\label{fig:Cdbe2} . Experimental and calculated $BE2$ rates
  for the Cd isotopes. \bet values from
  from~\cite{Cd100-104},\cite{Cd102-104}, and ~\cite{be22016}. \bef
  values from~\cite{nudat2}. In parenthesis $(e_{\nu})$, $e_{\pi}=1.40$ is fixed}
\end{figure}

In Fig.~\ref{fig:Cdbe2} it is seen that $utM=000$ and 101 give the
same results provided $e_{\nu}$ is properly chosen. There is little
difference between $utM=111$ and $utM=101$ because as soon as neutrons
are added they block the corresponding jumps, as mentioned above.
 
The calculation yields near perfect agreement with the Alaga rule:(
\beq=$\approx 1.43$\bet). In the figure it is shown for $utM=101$ but
it holds as well for 000 and 111. The more stringent quadrupole
quotient rule Eq.~\eqref{qQ} yields an average $Q/q=0.96$ for
\A{106-110}Cd, corroborating the existence of a deformed region.
A possibility anticipated in Ref.~\cite{PhysRevC.96.014302}.

\section{The Sn isotopes}\label{sec:sn}

In moving to the tin isotopes the P$r_4$ part of the SP scheme becomes
isolated and sensitive to details of the interaction. B\"ack and
coworkers~\cite[Fig. 3]{BE2Sn100} suggest that a parabolic trend as
found in Banu \etal~\cite{banu2005}, or schematically in
Fig.~\ref{fig:SnBE2}, can be modified in a $utM=000$ context, by
changes in the single-particle behavior, thus leading to the first
tentative explanation of the plateau. The more complete calculations
of Togashi \etal~\cite[Fig. 2]{PhysRevLett.121.062501} demand $g$
excitations to achieve a satisfactory result, very close to ours in
the upper Fig.~\ref{fig:be2042}, in spite of huge differences in the
$g$ proton occupancies (spin and mass dependent in their case and
nearly constant in ours). No \bef results are proposed in this
reference.

\subsection{The \beq anomaly and the pairing-quadrupole
    interplay}

\begin{figure}[htb]
  \begin{center}
\includegraphics[width=0.4\textwidth]{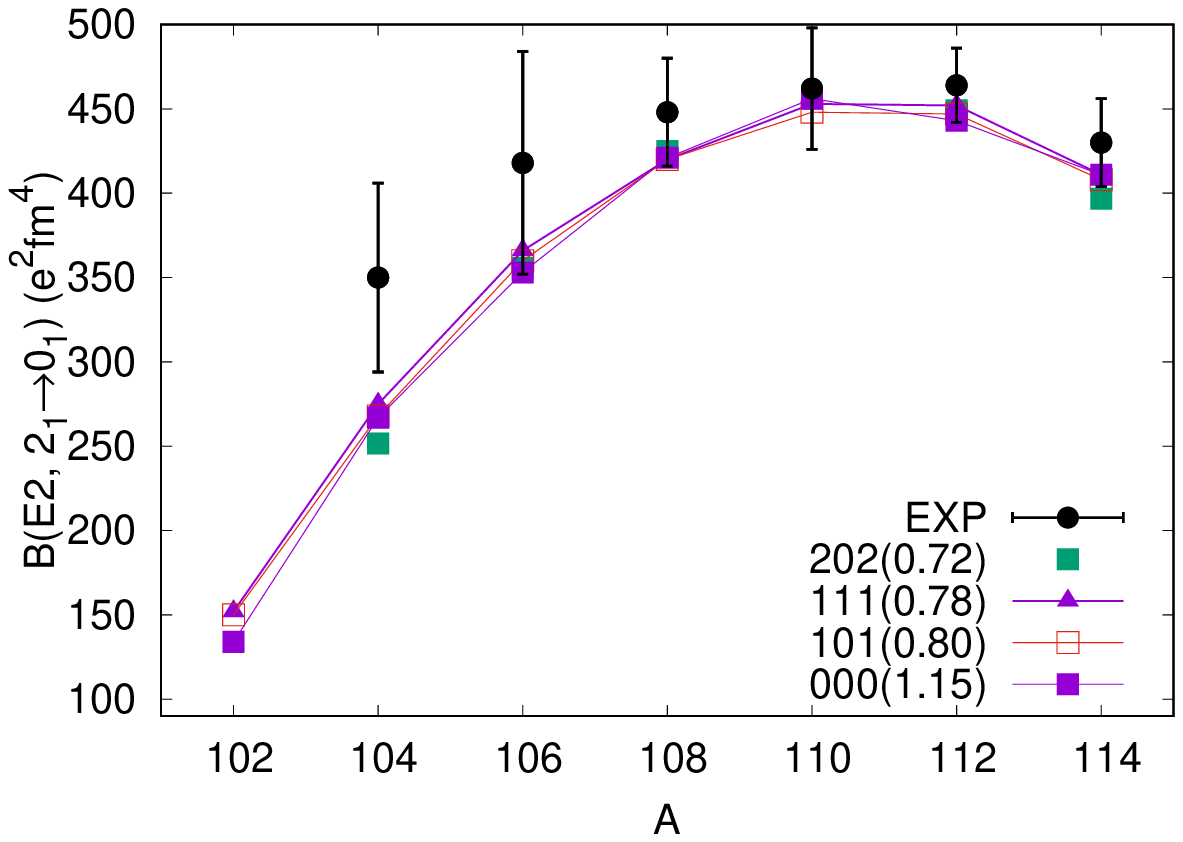}
\includegraphics[width=0.4\textwidth]{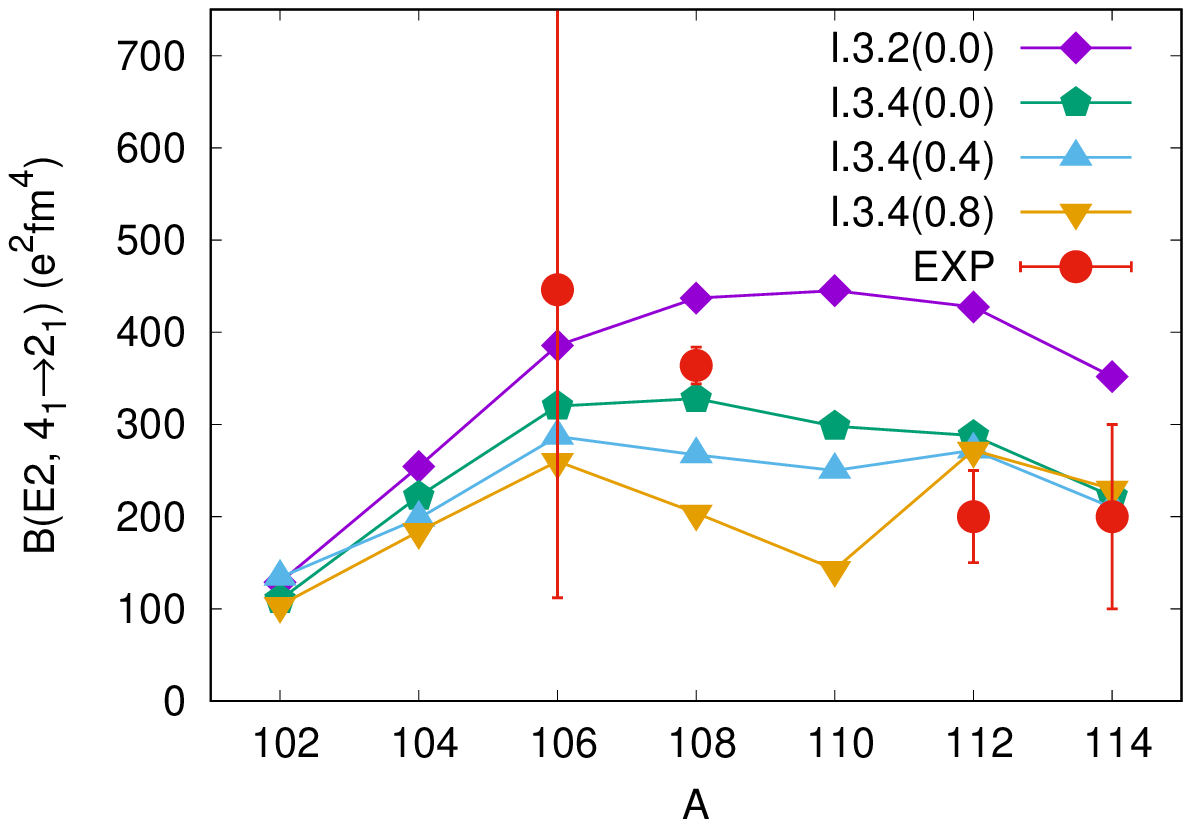}
\end{center}
\caption{\label{fig:be2042}.Upper panel: Experimental and calculated
  \bet values for the Sn isotopes. I.3.4 interaction. In parenthesis
  $(e_{\nu})$, $e_{\pi}=1.40$ is fixed. Experimental values
  from Ref~\cite{be22016}. Lower panel \bef data from Jonsson
  \etal~\cite{JONSSON1981} for \A{112-114}Sn, and from Siciliano
  \etal~\cite{marco} for \A{106-108}Sn. I.3.2($\delta$) and
  I.3.4($\delta$) calculations with the $s_{1/2}$
  single-particle energy displaced by $\delta=$0.0, 0.4 and 0.8 MeV
  with respect to the GEMO value of 0.8 MeV.}
\end{figure}

  The basic tenet of this paper is that quadrupole dominance is
  responsible for the \bet patterns in the light Cd and Sn
  isotopes. Which means that they should exhibit a pseudo SU3
  symmetry. Hence, we expect the existence of an intrinsic state,
  implying the validity of the Alaga rule (defined after
  Eq.~\eqref{bme2}). The expectation is fulfilled in Cd
  (Fig.~\ref{fig:Cdbe2}) but it fails in Sn, as seen in
  Fig.~\ref{fig:be2042}. Let us examine what to make of it.

  The \bet rates are consistent with pseudo SU3 validity, and are
  inmune to details. The \bef rates are sensitive to the
  single-particle field and to the pairing strength

  To test the influence of the single particle field on \bef rates,
  the energy of the $s_{1/2}$ orbit in \A{101}Sn was displaced by 0.0,
  0.4 and 0.8 MeV with respect to the present GEMO choice~\cite{gemo},
  called DZ (Duflo Zuker) in Ref.~\cite[Fig. 3.2.1]{Sn100} where an
  extapolated value (EX) is given as reference. The position of the
  $s_{1/2}$ orbit for DZ and EX differ by 800 keV. In the calculations
  reported In Fig.~\ref{fig:be2042}, I.3.4(0,0) and I.3.4(0.8)
  correspond to DZ and EX respectively.  The \bef differences are
  significant. Thanks to the recent \A{108}Sn \bef measure of
  Siciliano \etal~\cite[Fig. 3b]{marco}, the DZ choice is clearly
  favored.

  The \beq$<1$ anomaly had been detected in
  \A{114}Xe~\cite{deangelis2002coherent}, in
  \A{114}Te~\cite{moller2005e2} , and more recently in \A{172}Pt,
  Ref.~\cite{cederwall2018}, where it is stressed that no theoretical
  explanation is available.  Here, the sensitivity to the pairing
    strength provides a clue. In Fig.~\ref{fig:be2042}, its decrease
    in going from I.3.4 to I.3.2 produces a substantial increase of
    \bef. As can be gathered from Ref.~\cite[Figs. 3a and 3b]{marco},
    \bet is totally inmune to pairing, while \bef is so sensitive that
    a sufficient decrease in strength could bring \beq close to the
    Alaga rule. It appears that pairing is eroding the deformed
    band. Only the lowest $J=0$ and 2 are spared, giving way to a
    pairing-quadrupole interplay, that will eventually end up in
    pairing dominance at $N\approx 70$. Next we examine how the
    transition may take place.

\section{Cd and Sn at ${\bm N \geq 64}$}

In Table~\ref{tab:Q2g} the naive P$r_4$ adimensional intrinsic
quadrupole moments for prolate (q0p) and oblate (q0o) are compared for
Sn. The former are the same as $q(n)_n$ in Table.~\ref{tab:SP}. The
latter are obtained by filling the platforms in reverse order (from
the top). Up to $N=56$ prolate dominates. From $N=58$ to 62 there is
oblate-prolate degeneracy. At $N=64$, oblate dominates. The trend in
sign indicated by the intrinsic values is respected by the calculated
spectroscopic moments that opt for ``oblate'' shapes for $N>58$ \ie
$A>108$. For \A{112-114}Sn the shell model results are close---for the
quadrupole moments---or agree---for the magnetic moments---with the
measured values. Note: The magnetic moments are very sensitive to the
anomalous $g_{l\nu}$ taken from Ref.~\cite{neutron-gl}.

\begin{table}[h]
  \caption{\label{tab:Q2g}Intrinsic adimensional q0 for prolate (q0p),
    and oblate (-q0o) states. Calculated spectroscopic quadrupole
    moments and g-factor, Q2, Q4, g for I.3.4 \enupi=0.72, 1.40;
    $g_{s\nu}$=-2.869, $g_{l\nu}$=-0.070~\cite{neutron-gl},
    $g_{s\pi}$=4.189,  $g_{l\pi}$=1.100. Experimental Q2* and g* from
    Allmond \etal~\cite{PhysRevC.92.041303}, $g_{s\nu\pi}$ quenched by  
    0.75 with respect to the bare values~\cite[Fig. 28]{rmp}} 
\begin{tabular*}{\linewidth}{@{\extracolsep{\fill}}|c|ccccccc|}
\hline
N &  q0p&    -q0o&         Q2&  Q4&        Q2*&      g*   &    g  \\
52&   12&       6&        -18& -24&	      &           & -0.157\\
54&   18&	12&       -21& -21&	      &           &  0.012\\
56&   24&	18&       -16& -17&	      &           &  0.103\\
58&   24&	24&        -5& -02&	      &           &  0.142\\
60&   24&	24&         3&  10&	      &           &  0.142\\
62&   24&	24&        14&  26&	  4(9)&  0.150(43)&  0.135\\
64&   18&	24&        25&  43&       9(8)&  0.138(63)&  0.106\\
\hline
\end{tabular*}
\end{table}

So far the $sdg$ space has proven sufficient as the effects of the
$h_{11/2}$ orbit ($h$ for short) remain perturbative. For Sn we know
from classic $(p,d)$ work~\cite{CAVANAGH197097}, that the $h$
occupancy---very small up to \A{110}Sn--- increases at \A{112-114}Sn,
as borne out by calculations that indicate the need of a boost of some
10\% in \bet~\cite{fred} with respect to Fig.~\ref{fig:be2042}. Beyond
$N=64$, the explicit inclusion of the $h$ orbit becomes imperative but
the situation is different for the two families, as can be gathered in
the transitional nuclei \A{112}Cd and \A{116}Sn.

For Cadmium the calculations give systematically prolate values in
line with Stone's Q tables~\cite{STONE2016}, but in \A{112}Cd
(excluded from both Table~\ref{tab:SP} and Fig.~\ref{fig:Cdbe2}) they
yield severe underestimates whose correction necessitates the
\tcn{introduction of a quasi-SU3 mechanism (referred generically as Q
  in what follows, and Q$hfh$ for the case we introduce next)}.

\begin{figure}[t]
  \begin{center}
\includegraphics[width=0.5\textwidth]{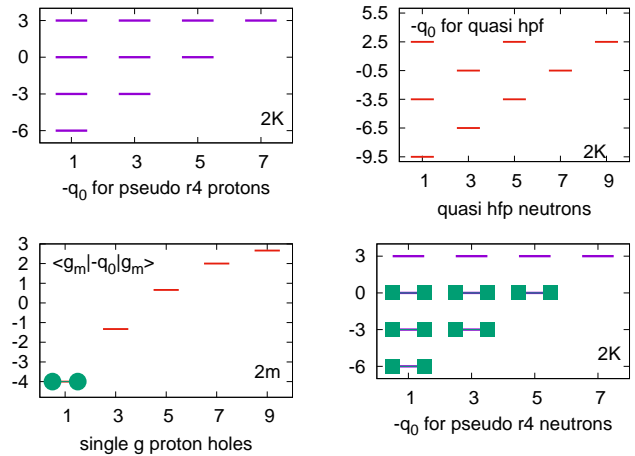}
\end{center}
\caption{\label{fig:Cdx} \tcn{The \A{110}Cd intrinsic state in the SPQ
    space. The few lowest quasi SU3 (Q$hfp$) $q_0$ platforms are obtained by
    diagonalizing the quadrupole operator $\hat{q_0}$ in the $hfp$
    space \ie the degenerate $\Delta J=2$ sequence in $pfh$ shell:
    $h_{11/2}, \, f_{7/2}, \, p_{3/2}$. A more realistic estimate must
    account for the splittings between the orbits which involves
    solving selconsistently Eq~\eqref{hfin} as explained in the
    text. The lowest resulting platforms are at \q=5.0, 5.0 (both
    $k=1/2$) and 4.0 ($k=3/2$).}}
\end{figure}

It is illustrated in Fig.~\ref{fig:Cdx}, which will serve as a basis
  to explore possible \bet. Starting from Eqs.(2,5), adapted  to
  \A{110+k}Cd (or \A{114+k}Sn, dropping the $8e_{\pi}$ term)

\begin{gather}\label{cd110be2}
B(E2: 2_1\to 0_1) =[8e_{\pi} +(24+\zeta)e_{\nu})b^2]^2/50.3 
\end{gather}
in \eff, which, using values from Table I, $e_{\nu}=1.1$ and $b^2=5$, yields
\bet=800\eff (use round numbers in all that follows) for $\zeta=0$,
\ie \A{110}Cd.
 
For \A{112}Cd addding a P$r_4$ pair ($\zeta=-6$) yields 600\eff, too
small against the observed 1000\eff.  The alternative is to promote
one or two Q$hfp$ pairs. As the $q_0$ platforms in the figure are
obtained under the assumption of single-particle degeneracy they yield
huge rates ($\zeta=19$ or 32) for a total 1800 and 2800\eff
respectively. A more realistic estimate demands that $q_0$ be
evaluated in the presence of a single-particle field.
  
To do so examine~\cite[Eq.(19)]{nilssonSU3} (remember $q_0=$\q)
\begin{gather}
H=H_{sp}-\frac{\hbar \omega \delta}{3}\hat{q_0}\equiv H_{sp}-\beta\hbar \omega \kappa
\frac{\hat {q_0}}{{\cal N}^2}q_0\label{hfin}\\
{\cal N}^2=\sum (2q_{20rs})^2=\sum_{k=0}^p (k+1)(2p-3k)^2\label{N2}.
\end{gather}

Eq.~\eqref{hfin} compares the classic Nilsson problem to the left and the
selfconsistent version to the right, which demands the solution of a
linearized $\hat{q_0} \hat{q_0}$ problem, subject to the condition that input and
output $q_0$ coincide. Hence, $q_0$ the quantity we are
after, is calculated while in the Nilsson case it is simply the
parameter $\delta$. To have all quantities entering the
calculation fully defined, the norm of the  $\hat{q_0} \hat{q_0}$
Hamiltonian~\cite[Eq.(12)]{nilssonSU3} is included: ${\cal N}^2$=420
for $p=5$.

Written in full, the $q_0$ part of $H$ involves the $S,\, P, \,Q$
cumulated pair contributions (the former from Eqs.(1,2)). Upon linearization, the
  factor $\beta$ ensures that the single $q_0=q_0(Q)/2$ couples to the full
  $q_0$, and not only to itself, \ie
  $\beta \langle q_0(Q)/2\rangle=\langle
  (q_0(S)/2+q_0(P)/2+q_0(Q)/2)\rangle$.
  Then $\beta=5$, since $q_0(S)/2=4, \, q_0(P)/2=12$ and
  $\langle q_0\rangle\equiv q_0(Q)=5.0$.

The single-particle energies are taken from GEMO~\cite{gemo}

 $\epsilon(h_{11/2})=0,\, \epsilon(f_{7/2})=2.0, \,
\epsilon(p_{3/2})=3.0$   MeV

Recapitulating. \hw=8.34, $\kappa=0.3$ (the same as in the I.3.x
  interactions), $\beta=5$, ${\cal N}^2$=420, $\langle q_0\rangle=5.0$.

Then, solving for the right hand side of Eq.~\eqref{hfin}, leads to
two $k=1/2$ platforms at \q=5 and the $k=3/2$ one at \q=4, (as
mentioned in Fig.~\ref{fig:Cdx}). Combining them, the following values
of $\zeta=-6,\,10,\,18,\,28$ lead, through Eq.(7), to
\bet=600,1300,1800,2500 \eff or, 19, 41, 56, 78 W.u. The first number is
obtained by adding a P$r_4$ pair, the next three by adding 1, 2 and 3
Q$hfp$ pairs rspectively. Note that for \A{114}Cd, the promotion of
two and three Q$hfp$ pairs can be combined with one or two \q=0
P{$r_4$} holes. As a consequence the spectrum may contain bands for
the four values above.  Experimentally all these bands (and a couple
more) are observed and fairly well reproduced by sophisticated
beyond-mean-field calculations~\cite{Cdgato}.  Here we only note that,
experimentally, the ground state transition
$B(E2: 2_1\to 0_1)=1000(100)$ \eff or 31(3) W.u, compares with
Rodr\'iguez's~\cite{Cdgato} 38 W.u, and our 41 W.u, which may be
somewhat reduced by increasing the $hfp$ single-particle splittings.

It is seen that in the Cd family the full panoply of quadrupole agents
{\bf SPQR} (R for representaions) is at play and can provide
qualitative and some useful semiquantitative results. Though only
\A{112}Cd has been examined, the context makes it clear that heavier
isotopes will remain well deformed and that $N=64$ signals a smooth
merger between an S$g$P$r_4$ scheme into a  S$g$P$r_4$Q$hfp$ one. 
That the heavier Cd isotopes are deformed was anticipated in
Ref.~\cite{Rodríguez2008}.

In the case of Sn, the interaction favors oblate at $N=64$ when it
becomes the only option (consistent with data in Allmond \etal
\cite{PhysRevC.92.041303}). If we were to pursue the analogy with the
Cd case we would expect the P$r_4$ scheme to merge into a P$r_4$S$h$
one. According to Eq.(1) the lowest contributions -\q= 5, 2.27 lead to
$\zeta=10,\,15$ and \bet= 700, 900 \eff= 22, 29 W.u. Experimentally
\bet=400 \eff= 12 W.u, too small, but there is a
$B(E2:0_2\to 2_1\approx$ 18 W.u (in both \A{116-118}Sn), suggesting
that the P$r_4$S$h$ strength splits about equally with some
intruder(s), as can be expected in a transition region. To go further,
shell model calculations are called for. Though unmanageable for Cd,
they are quite feasible in neutron only spaces throughout the Sn
region.  As they explain \bet patterns in \A{102-114}Sn and
\A{120-132}Sn, it is likely they will also do so in
\A{116-120}Sn provided I.3.4 is adjusted to include
the $h_{11/2}$ orbit, and a more sophisticated monopole treatment of
GEMO.

\section{The interaction and the model spaces}\label{sec:SM}

We have just mentioned the consequential result emerging from Fig. 4:
the possibility to describe the \bet pattern through a neutron-only
calculation (the 000 case). This is at variance with previous
calculations ~\cite{banu2005,ekstrom2008sn,BE2Sn100} using the CDB
(Charge Dependent Bonn) potential~\cite{CDB}, renormalized following
Ref.~\cite{Hjorth-Jensen.Kuo.Osnes:1995}. Which raises two questions:
why our I.3.x interactions succeeds where others fail?  and why the
neutron-only description is viable? They can answered simltaneously
and we start by explaining how severely truncated spaces may represent
the exact results, by comparing the largest calculation available with
smaller ones. In Reference ~\cite{marco}, results are given for
\A{106-108}Sn in $utM=444$ ($m$-dimensions $10^{10}$) using the same
interaction (called B in what follows) as in Banu
\etal~\cite{banu2005} (but omitting the $h_{11/2}$ orbit).
\begin{figure}[h]
  \begin{center}
\includegraphics[width=0.4\textwidth]{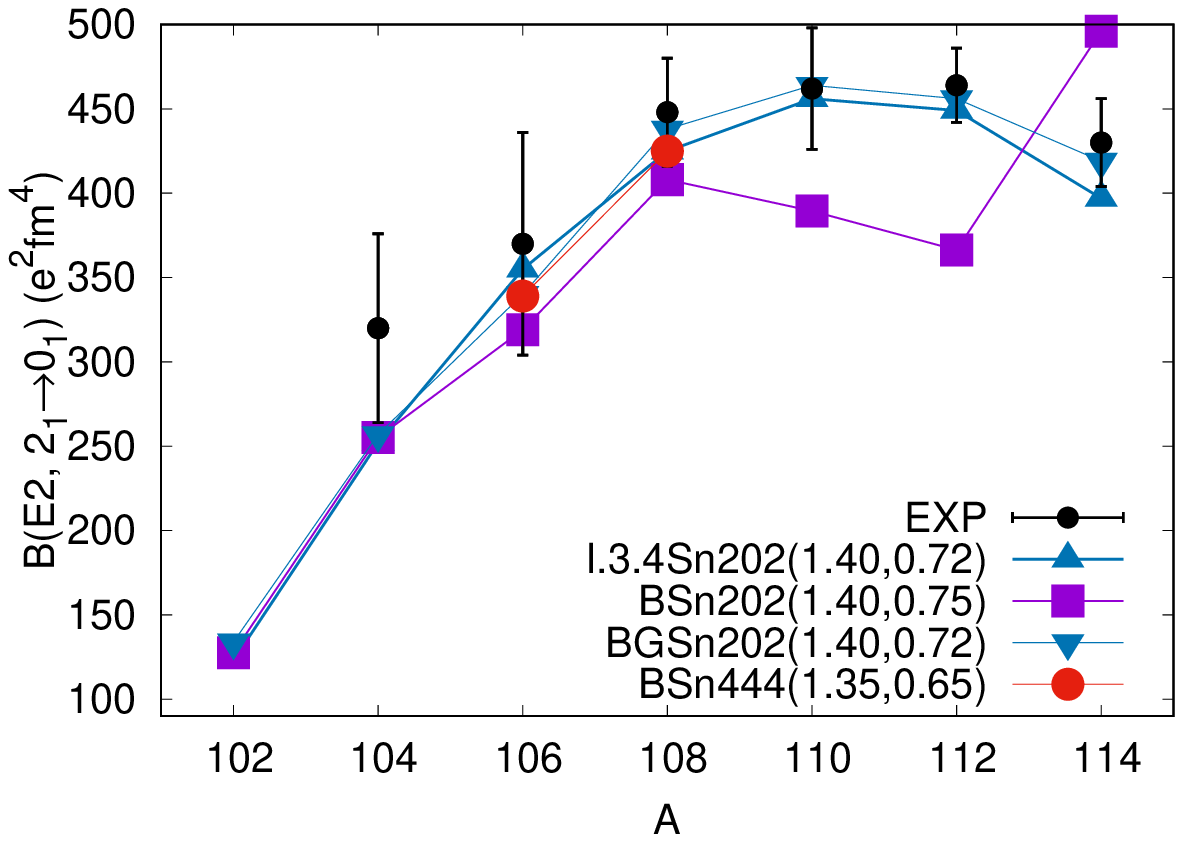}
\includegraphics[width=0.4\textwidth]{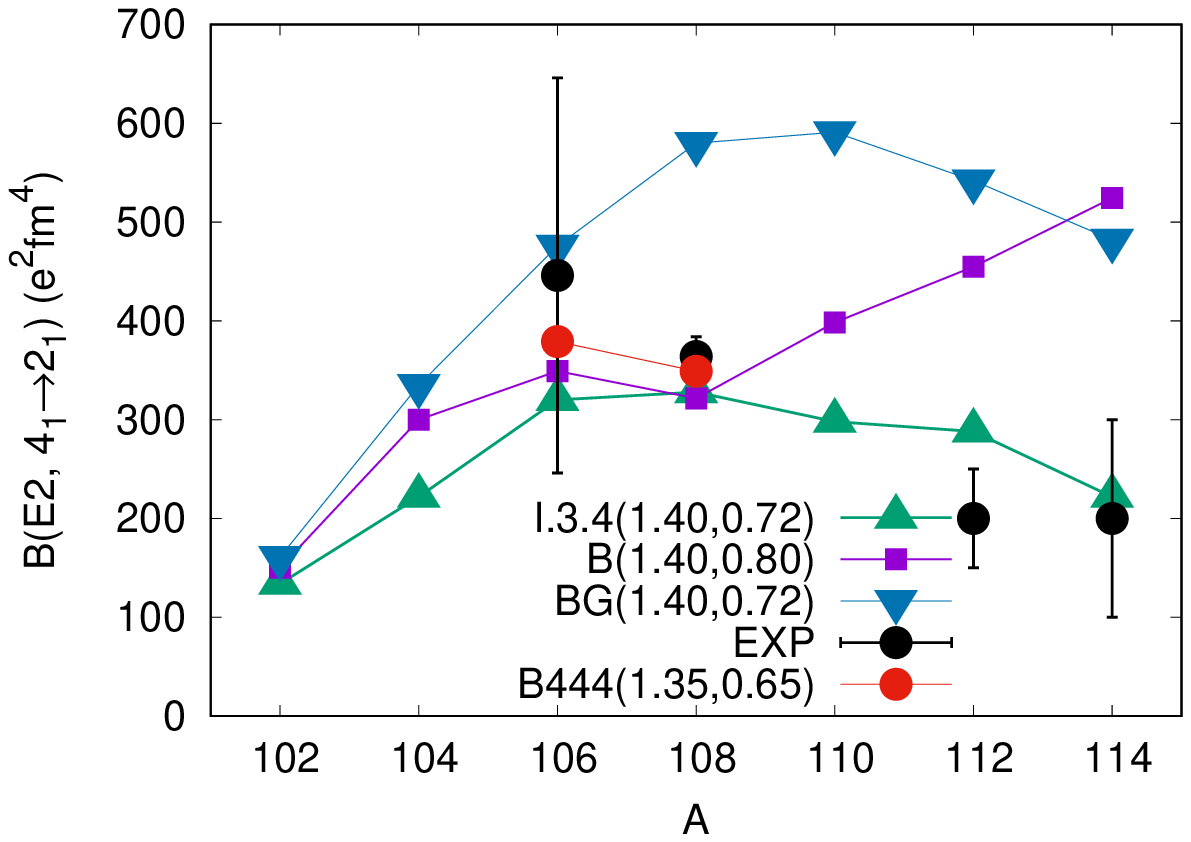}
\end{center}
\caption{\label{fig:comp}Comparing calculations for B and I.3.4
  interactions in the Sn isotopes. In parenthesis \epinu. Experimental
  data as in Fig.~\ref{fig:be2042}. See text }.
\end{figure}
In Fig.~\ref{fig:comp} it is shown as B444 (circles) and compared
with B202 (squares, the same interaction in our standard space).
The agreement is very good for the two points in \bet and \bef. The result
amounts to a splendid vindication of the shell model viewed as the
possibility to describe in a small space the behavior of a large
one. Although in general the reduction from large to small spaces
demands renormalization of the operators involved, for our purpose only
the effective charges are affected. A non trivial fact that invites
further study.

For much of the region, discrepancies between I.3.4 and B can be traced
to poor monopole behavior of the latter. If the interaction is made
monopole free and supplemented by the GEMO single-particle field used
in our I.p.q forces, the resulting BG202 in Fig.~\ref{fig:comp}
produces \bet patterns identical to the ones for I.3.4-202, while for
\beq the pattern is close to I.3.0 which is not shown, but can be
guessed by extrapolation in Fig.~\ref{fig:be2042} and from the
analysis in Ref.~\cite{mdz} revealing the same $q\cdot q$ content in
I.3.4 and B, and a much weaker pairing for the latter; so weak in fact,
that the B results come close to the Alaga rule.

It follows that for I.3.0, say, the P$r_4$ symmetry will hold, at
least partially. As the pairing force is switched on, the $J=0_1,2_1$
states are not affected, while $J=4_1$ is. Which points to an unusual
form of interplay between the two coupling schemes---pairing and
quadrupole---traditionally associated to collectivity. In single fluid
species, such as Sn, the seniority scheme can operate fully. It breaks
down in the presence of two kind of particles, which turns to be the
condition for quadrupole to operate successfully, as indicated by the
Cd isotopes. What is unusual is the presence of quadrupole coherence
in the light tins. It is shaky and challenged by pairing and we know
that at about $A=120$ the seniority scheme will prevail. For the
transition nuclei \A{116-118}Sn, mixing of spherical and weakly oblate
states is expected. Neutron-only calculations---in which the monopole
field will play a crucial role---are likely to shed light on these
matters.

The Sn isotopes were an example of pairing, then of quadrupole, then of
pairing again. The world is but a perennial swing (Le monde n'est
qu'une branloire perenne. Essais III 2~\cite[p. 584]{montaigne}).
\begin{acknowledgments}

Alfredo Poves and Fr\'ed\'eric Nowacki took an active interest in the
paper and made important suggestions.The collaboration with Marco
Siciliano, Alain Goasduff and Jos\'e Javier Valiente Dob\'on is
gratefully acknowledged.
\end{acknowledgments}

\bibliography{ZukerS.bib}
\end{document}